\documentclass[12pt]{emulateapj}
\usepackage{natbib}
\usepackage{epsfig}

\shorttitle{A Extreme Water Maser Outflow} \shortauthors{Titmarsh et al.}

\newcommand{\etal}{et~al.} 

\newcommand{\UCHII}{UCH{\sc ii} }

\newcommand{\kms}{$\mbox{km~s}^{\scriptsize{-1}}$}
\newcommand{\msol}{\mbox{M\hbox{$_\odot$}}}
\newcommand{\lsol}{\mbox{L\hbox{$_\odot$}}}
\newcommand{\lta}{\raisebox{-0.6ex}{$\,\stackrel
{\raisebox{-.2ex}{$\textstyle <$}}{\sim}\,$}}
\newcommand{\gta}{\raisebox{-0.6ex}{$\,\stackrel
{\raisebox{-.2ex}{$\textstyle >$}}{\sim}\,$}}   

\newcommand{\pasa}{PASA}

\begin{document}


\title {G\,10.472+0.027: An Extreme water maser outflow associated with a Massive Protostellar Cluster}
\author
 {A.M.~Titmarsh\altaffilmark{1, 2}, S.P.~Ellingsen\altaffilmark{1}, S.L.~Breen\altaffilmark{2},
J.L.~Caswell\altaffilmark{2}, M.A.~Voronkov\altaffilmark{2}}

\altaffiltext{1} {School of Mathematics and Physics, University of
Tasmania, Hobart, Tasmania, Australia}
\altaffiltext{2} {CSIRO Astronomy and Space Science, Australia Telescope National Facility, PO Box 76
Epping, NSW 1710, Australia}

\label{firstpage}


\begin{abstract}

An Australia Telescope Compact Array search for 22~GHz water masers towards 6.7~GHz class~II methanol masers detected in the Methanol Multibeam (MMB) survey has resulted in the detection of extremely high velocity emission from one of the sources.  The water maser emission associated with this young stellar object covers a velocity span of nearly 300~\kms.  The highest velocity water maser emission is red-shifted from the systemic velocity by 250~\kms, which is a new record for high-mass star formation regions.  The maser is associated with a very young late O, or early B star, which may still be actively accreting matter (and driving the extreme outflow).  If that is the case future observations of the kinematics of this water maser will provide a unique probe of accretion processes in the highest mass young stellar objects and test models of water maser formation.

\end{abstract}


\section{Introduction}

Maser emission from the 22~GHz transition of water is commonly observed towards a number of different types of astrophysical objects including high-mass star formation regions \citep[e.g][]{Genzel+77}, late-type stars \citep[e.g.][]{Engels79} and the nuclei of active galaxies \citep[e.g.][]{Claussen+84}.   Water masers appear to be the most common of all interstellar maser species \cite[e.g.][]{Breen+11b,Walsh+11} and there are thought to be many thousands of sources in the Milky Way alone.

Water masers are collisionally pumped, and towards high-mass young stellar objects they arise in the high density (10$^9$ cm$^{-3}$), warm (400 K) post-shock gas produced in molecular outflows and highly collimated jets  \citep{Elitzur+89}. When the direction of the outflow is approximately perpendicular to the line of sight we observe strong water maser emission at velocities close to the systemic velocity, as this orientation produces longer velocity coherent lines of sight through the post-shocked gas with only a small component of the outflow velocity directed along the line of sight \citep{Genzel+81a}.  When the motion of the masing gas is close to the line of sight of the observer, emission significantly offset from the systemic velocity of the region is observed, but it is typically weaker than emission at the systemic velocity due to less favourable geometry.  Under this scenario the 3D velocity of the high- and low-velocity water maser emission is similar and it is the geometry of the source and its effect on the path length of the maser which produces the observed spectrum of the water maser \citep[e.g.][]{Hollenbach+13}.  Where observations of the proper motion of water masers have been made \citep[e.g.][]{Genzel+81a} they are broadly consistent with this picture, and show a tendency for masers at close to the systemic velocity to exhibit larger proper motions than any high-velocity emission in the same source.

The details of the process through which high-mass stars form remains controversial, with monolithic collapse \citep[][]{Krumholz+09} and competitive accretion \citep[e.g.][]{Bonnell+06} being the two currently favoured alternatives.  In the monolithic collapse model the asymmetries produced by disks and jets are the means by which radiation pressure is overcome in some parts of system, allowing accretion to continue \citep{Krumholz+09}.  \citet{Beuther+05} suggest that the most collimated outflows are associated with the youngest sources and it is well established that the higher the source luminosity the more energetic the outflow \citep[e.g.][]{Shepherd+96a}.  So the expectation is that the most highly collimated, energetic outflows are associated with the highest mass young sources.  To date there are no well established examples of accretion disks associated with proto-O stars, the best estimates for the mass of young stellar objects with clear signs of active accretion are that they are proto-B stars \citep[e.g. IRAS20126+4104,][]{Cesaroni+05}.  In searching for the highest mass accreting protostars, identifying those sources associated with the highest velocity outflows may be the most efficient means.  

Water masers are known to commonly show high-velocity emission, offset from the systemic velocity of the source, sometimes in excess of 100~\kms\/ \citep{Breen+10b}.  So searching for the water maser sources with the most extreme high-velocity emission in a large sample is an obvious strategy to identify candidate proto-O stars.  6.7 GHz methanol masers are known to exclusively trace high-mass star formation \citep{Minier+03,Breen+13b} and to trace a very early phase of the high-mass star formation process \citep[e.g.][]{Ellingsen06}.  We are undertaking sensitive, high-spatial and spectral resolution observations for water maser emission towards all 6.7~GHz methanol masers detected in the Parkes Methanol Multibeam (MMB) survey \citep{Green+09a}.  The MMB is a sensitive, unbiased search for 6.7~GHz methanol masers covering the entire southern Galactic Plane.  To date the results covering the longitude range $l = 186^{\circ}$--$20^{\circ}$ have been published \citep{Caswell+10,Caswell+11,Green+10,Green+12a}.  The general results of the followup water maser survey will be presented in a series of upcoming papers (Titmarsh \etal\/ 2013, in preparation); here we discuss the water maser with the most extreme velocity range of any high-mass star formation region detected as part of these observations.

\section{Observations}

The observations of the water maser emission towards G\,10.472+0.027 (6.7-GHz methanol maser position $\alpha_{J2000}$ 18h 08m 38.20s; $\delta_{J2000}$ -19$^{\circ}$ 51$^{\prime}$ 50.1$^{\prime\prime}$) were undertaken with the Australia Telescope Compact Array (ATCA) on 2010 November 3.  The array was in the H214 configuration.  The synthesised beam width for the observations was approximately 11.1$^{\prime\prime} \times$ 8$^{\prime\prime}$, however, the astrometric accuracy (which is primarily determined by the atmospheric conditions and the properties of the phase calibrator) is better than 2$^{\prime\prime}$ for these observations \citep{Breen+10b}.  The Compact Array Broadband Backend (CABB) was used for the observations \citep{Wilson+11} and configured with a 64 MHz bandwidth with 2048 spectral channels centred on the rest frequency of the 22~GHz water maser transition.  This bandwidth gives a total velocity coverage $>$ 800~\kms\/ and velocity resolution of 0.51~\kms\/ for uniform weighting of the correlation function.

The data were reduced with {\sc miriad}, using the standard techniques for ATCA spectral line observations.  Amplitude calibration was with respect to PKS\,B1934$-$638, for which we assumed a flux density of 0.81~Jy at a frequency of 22.235~GHz, and PKS\,B1253$-$055 was the bandpass calibrator.  The observing strategy interleaved 90 second observations of each of a group of six target 6.7 GHz methanol maser sources (such as G\,10.472+0.027) with 90 second observations of the phase calibrator 1730-30 before and after the target source block.  The total on source time for the observation of  G\,10.472+0.027 was 7.2 minutes, spread over an hour angle range of 7 hours.  The resulting RMS noise in a single 0.42~\kms\/ spectral channel image was 20~mJy beam$^{-1}$.  The results for the observations of the other MMB targets in the Galactic Longitude range $l$ = 6--20$^{\circ}$ will be reported in Titmarsh et al. (2013, in preparation).

Additional ATCA observations were made on 2013 June 06 in a Director's time allocation.  For these observations the array was in the 6C configuration and only 4 of the 6 antennas were available.  The observations consisted of four 10 minute scans on the target source, interleaved with 2 minute observations of the phase calibrator.  The same primary flux, bandpass and phase calibration sources were used as for the 2010 observations.  The total hour-angle range for the observations was less than 1 hour, so with an East-West aligned array the data was not suitable for imaging; however, we were able to extract a sensitive spectrum from the {\em uv}-data by vector averaging at the location of the maser emission (as the maser emission is a point source to the array all the flux is recovered).

\section{Results and Discussion}

\begin{figure*}
\begin{center}
\includegraphics[width=8cm,angle=-90]{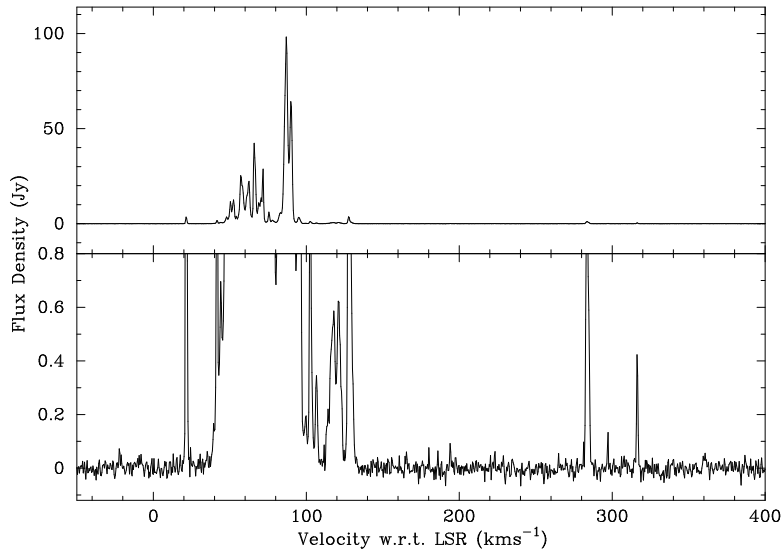}
\end{center}
\caption{The 22 GHz water maser emission associated with G\,10.472+0.027 observed on 2010 November 3 extracted from the image cube formed after phase-only self-calibration of the data using a 1 minute solution interval and a model based on the strongest water maser emission at 87~\kms.  The bottom panel shows the extent of the weak, high velocity emission.}
\label{fig:spectrum}
\end{figure*}

Our spectra of G\,10.472+0.027 are shown in Figs.~\ref{fig:spectrum} \& \ref{fig:compare} (see Section~\ref{sec:detection} for a discussion of Fig.~\ref{fig:compare}).  The position of the water maser emission (measured from an image produced prior to self-calibration) is $\alpha_{J2000}$ = 18$h$ 08$m$ 38.24$s$, $\delta_{J2000}$ = -19$^{\circ}$ 51$^{\prime}$ 50.2$^{\prime\prime}$.    The bottom panel of Figure~\ref{fig:spectrum} shows that the total velocity range for this source extends from +21 to +317 \kms, a total span of nearly 300~\kms\/.  Our observations cover the LSR velocity range $-$420 to +402 \kms\/ (i.e. a significantly larger extent than shown in Figure~\ref{fig:spectrum}).  We also imaged the data with a coarser 1~\kms\/ velocity resolution and are confident that we have detected all emission stronger than $\sim$50 mJy over the observed velocity range.  We have used the self-calibrated data to compare the measured location of the strongest systemic water maser emission with the the high velocity features and measure an offset of $<$ 0.1 $^{\prime\prime}$ (i.e. the same position to within the astrometric uncertainty).  

Water maser emission significantly offset from the systemic velocity of the region is not particularly unusual for  high-mass star formation regions; however, sources where the total velocity range significantly exceeds 100~\kms\/ are.  Velocity spans in excess of 300~\kms\/ have been observed towards several post-asympototic giant branch ``water fountain'' sources \citep[e.g. OH\,009.1-0.5 and IRAS 18113-2503][]{Walsh+09,Gomez+11}.  The 296~\kms\/ range we observe for G\,10.472+0.027 exceeds the velocity span observed for any water maser associated with a high-mass star formation region with the exception of W49N, for which \citet{Walker+82} measure a $\sim$400~\kms\/ range.  In W49N there are both highly red-shifted  and blue-shifted water maser features extending $\sim$170~\kms\/ and $\sim$230~\kms\/ from the systemic velocity, respectively \citep{Walker+82}.  Observations of a number of ammonia transitions towards G\,10.472+0.027 give a peak velocity of the emission in the range +66.8 to +68.8~\kms\/ \citep{Churchwell+90,Purcell+12}.  This means the highest velocity emission detected in our ATCA observations is red-shifted by 250~\kms, which exceeds the previous greatest offset from the systemic velocity observed in the blue-shifted water maser emission in W49N.  Recent theoretical modelling has demonstrated that water masers can be produced by dissociative J shocks with speeds ranging from 30 to 200~\kms\/ \citep{Hollenbach+13}.  Our observations suggest that some water masers may be associated with even higher velocity shocks.  Very long baseline interferometry observations of the water masers in G\,10.472+0.027 which measure the brightness temperature and physical dimensions of the emission region will allow a direct test of this theory, as \citet{Hollenbach+13} specify equations which allow the calculation of critical model parameters directly from observable quantities.

\subsection{Previous observations of masers in the  G\,10.472+0.027 region}

Water maser emission associated with G\,10.472+0.027 was first detected more than 35 ago by \citet{Genzel+77} in a search towards star formation regions made with the Efflesberg 100m radio telescope.  The source is close (in terms of angular separation) to the well known W31 star formation region and was initially identified as being associated with it.  However, \citet{Green+10} argue that the very different systemic velocity of G\,10.472+0.027 ($\sim$70~\kms\/) compared to other sources in the W31 region (systemic velocities $<$ 40~\kms) suggests that they are two unassociated regions which happen to lie along similar lines of sight.  The results of the various water, OH and methanol maser observations that have been made towards G\,10.472+0.027 are summarised in Table~\ref{tab:masers}.  

\begin{table*}
  \scriptsize
  \begin{center}
  \caption{Previous maser observations towards G\,10.472+0.072}
  \begin{tabular}{lllllllll} \hline
  {\bf Maser}      & {\bf Telescope} & {\bf RA}         &  {\bf Dec.)}     & {\bf Date}          & {\bf Peak}  & {\bf Vel.}         & {\bf Peak}           & {\bf Ref.} \\
  {\bf Type}        &                             & {\bf (J2000)} &  {\bf (J2000)} &                           & {\bf Vel.}     & {\bf Range}   & {\bf Flux}      & \\ 
                           &                             &                        &                         &                           & \multicolumn{2}{c}{\bf (\kms)}     & {\bf (Jy)}       & \\ \hline
  22 GHz water & Effelsberg          & 18:08:38.40 &  -19:51:51.7  & 1976 Oct          & 64                & +20 -- +100         & 230            &  \citet{Genzel+77}  \\
  22 GHz water & Effelsberg          & 18:08:38.20 & -19:51:49.5   & 1989 Apr         & 61             & +35 -- +125  & 485            &  \citet{Churchwell+90} \\
  22 GHz water & VLA                     & 18:08:38.20 & -19:51:49.5   & 1991 Dec        & 64             & 84 (total)      & 202            &  \citet{Hofner+96} \\
  22 GHz water & ATCA                  & 18:08:38.30 & -19:51:48.8   & 2003 Oct         & 60                 & +30 -- +93            & 45              &  \citet{Breen+10b} \\
  22 GHz water  & ATCA                  &  ~~~~~~$\prime\prime$ & ~~~~~~$\prime\prime$ & 2004 Jul          & 62                 & +28 -- +129          & 169           &  ~~~~~~$\prime\prime$ \\
  22 GHz water & Mopra                 & 18:08:37.4  & -19:51:28       & 2007-2010     & 71             & +41 -- +88             & 108           &  \citet{Walsh+11} \\     
  22 GHz water & ATCA                  & 18:08:38.24 & -19:51:50.2   & 2010 Nov        & 87                & +21 -- +317          & 100            &  current work\\
  22 GHz water & ATCA                  &   ~~~~~~$\prime\prime$ & ~~~~~~$\prime\prime$ & 2013 Jun         & 62 & +30 -- +300         & 65 &  ~~~~~~$\prime\prime$ \\
  1.6 GHz OH   & ATCA                  & 18:08:38.25 & -19:51:49.4   & 1994 Nov        & 52            & +48 -- +69    & 1.2        &  \citet{Caswell98} \\
  1.6 GHz OH   & Parkes                &   ~~~~~~$\prime\prime$ & ~~~~~~$\prime\prime$ & 2004 Nov         & 52 & +44 -- +71         & 1.4 &  \citet{Caswell+13a} \\
  1.6 GHz OH   & Parkes                &   ~~~~~~$\prime\prime$ & ~~~~~~$\prime\prime$ & 2005 Oct         & 52 & +44 -- +71         & 1.5 &  ~~~~~~$\prime\prime$ \\
  6.7 GHz meth & ATCA                 & 18:08:38.20 & -19:51:50.1   & 1993 Feb         & 75               & +58 -- +77 & 120       &  \citet{Caswell+95c} \\
  6.7 GHz meth & Parkes               &  ~~~~~~$\prime\prime$ & ~~~~~~$\prime\prime$ & 2006-2009   & 75 & +58 -- +78         & 35 &  \citet{Green+10} \\
  12.2 GHz meth & Parkes             & ~~~~~~$\prime\prime$ & ~~~~~~$\prime\prime$ & 2008 Jun & 75      & +74 -- +77 & 12.4      &  \citet{Breen+10a} \\ \hline
  \end{tabular}
  \label{tab:masers}
  \end{center}
\end{table*}

The first interferometric observations of the water maser emission in this source were made by \citet{Hofner+96} using the VLA.  These observations, with a $\sim$ 0.5\arcsec\/ beam found three clusters of water maser emission elongated in a direction roughly north-south, with an angular extent of $\sim$ 2\arcsec. The central cluster contains the strongest water masers with the widest velocity spread and is projected against an \UCHII region.  The absolute position we measure for the water masers is coincident (to within the astrometric uncertainty) with the location of the OH and 6.7~GHz methanol masers and previous interferometric observations of the water maser emission in this region (see Table~\ref{tab:masers}).  Interestingly the velocity ranges of both the OH and 6.7 GHz methanol masers towards G\,10.472+0.027 exceed 20~\kms\/ and so are significantly larger than typically observed in these transitions \citep[e.g.][]{Caswell98,Caswell09}.  Another peculiarity of this region is that \citet{Cesaroni+98} detected blue-shifted absorption in NH$_3$(4,4) observations made with the VLA, indicating expansion in the thermal molecular gas.

\subsection{Detection of high-velocity water masers} \label{sec:detection}

\begin{figure}
\begin{center}
\includegraphics[width=6cm,angle=-90]{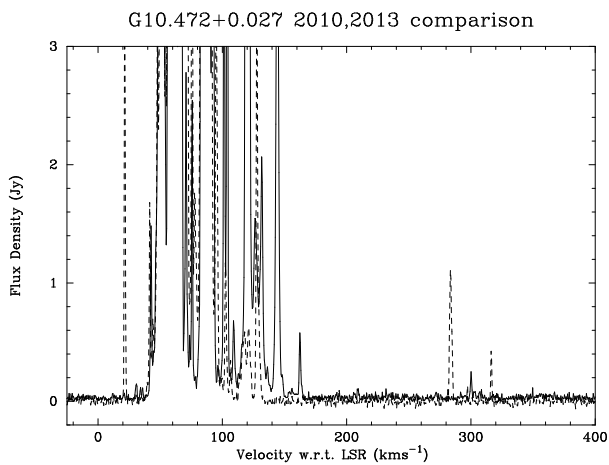}
\end{center}
\caption{A comparison of the water maser emission from G\,10.472+0.027 observed on 2010 November 3 (dashed line) and 2013 June 6 (solid line).}
\label{fig:compare}
\end{figure}

\citet{Caswell+10b} define a high velocity water maser feature to be one offset from the systemic velocity of the region by more than 30~\kms.  Using this criterion they find that 77 of 229 sources in \citet{Breen+10b} show one or more high velocity features (33\%).  Given that the G\,10.472+0.027 water masers have been observed on numerous occasions over a period of more than 35 years, a natural question to ask is why has the extreme velocity range never previously been noticed?  There appear to be two likely reasons.  The first is that most of the previous observations did not cover a sufficiently large
velocity range (and in a few cases had insufficient sensitivity).  The second possible reason as to why there has been no previous detection of high velocity emission in this source is that high velocity components are particularly variable \citep{Breen+10b}.  Comparing our ATCA observations of G\,10.472+0.027 made in 2010 November and 2013 June, the observed velocity range is similar (see Table~\ref{tab:masers}).  Figure~\ref{fig:compare} overlays the two spectra on a scale where the high velocity emission can be compared and shows that there is no emission \gta\/ 50 mJy observed in 2013 at the velocity of the 2010 highest velocity component.  However, although it is not clear from Fig.~\ref{fig:compare}, comparison of the emission near the systemic velocity between the two epochs also shows essentially no correspondence between individual spectral components between the two epochs.  So on the two occasions where observations with sufficient velocity range and sensitivity (RMS \lta 100 mJy) have been undertaken high-velocity emission has been detected, but is variable.  This suggests that future observations of known Galactic water masers with greater velocity coverage and high sensitivity will likely detect very high velocity emission in some of these sources as well.

\subsection{The Nature of the Star Formation Region}

We argued in the introduction that the presence of very high velocity water maser emission may be one of the most effective ways of finding a proto-O star, so what can be inferred by the different maser transitions and other observations that have been made toward G\,10.472+0.027?  The most critical parameter for any interpretation of the physical characteristics of the star formation region is the estimated distance.  This source has been observed as part of the BeSSeL project \citep{Reid+09} and the distance measured from trigonometric parallax measurements is 8.55 $\pm ^{0.62}_{0.55}$~kpc (Sanna et al., in preparation).  
 
The peak isotropic luminosity of the 6.7 and 12.2~GHz methanol masers in G\,10.472+0.027 are $\sim$2560 Jy\,kpc$^2$ (using the MMB flux density measurement) and $\sim$880 Jy\,kpc$^2$ respectively, placing it in the top 20\% of most luminous class~II methanol maser sources.  \citet{Breen+10a} and \citet{Ellingsen+11a} have proposed a maser-based evolutionary timeline for high-mass star formation regions.  On this basis \citet{Ellingsen+11a,Ellingsen+13a} would predict that this region is likely to exhibit emission from the 37.7~GHz methanol transition and perhaps other rare class~II masers.  The highest luminosity class II methanol maser emission is thought to occur close to the end of that evolutionary phase \citep{Breen+10a}, and the presence of a strong \UCHII region (integrated flux density $\sim $100 mJy at 22~GHz) and OH maser emission in this source is also consistent with that interpretation.  The presence of class~II methanol masers in high-mass star formation regions with ongoing accretion is supported by observations that suggest they are sometimes located within accretion disks \citep{Minier+98,Cesaroni+05} or infalling gas \citep{Goddi+11}.  If that is the case then we would expect the most evolved sources to also have generally higher mass.

\begin{figure}
\begin{center}
\includegraphics[width=9cm, angle=-90]{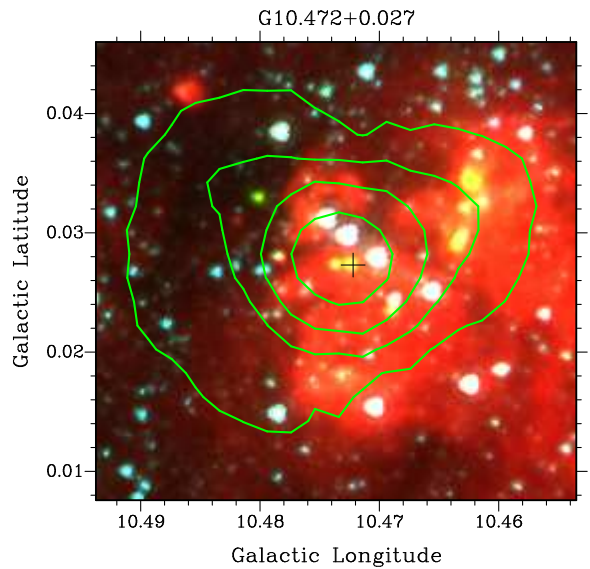}
\end{center}
\caption{A GLIMPSE 3 colour image of of G\,10.472+0.027 (red is 8.0~\micron, green is 4.5~\micron\/ and blue is 3.6~\micron), the green contours are the 1.1mm Bolocam Galactic Plane Survey data (at levels of 8, 16, 32 and 64\% of 22 Jy beam$^{-1}$, the peak intensity in the image).  The location of the water maser is marked with a black cross.}
\label{fig:IR}
\end{figure}

\begin{figure}
\begin{center}
\includegraphics[width=8cm]{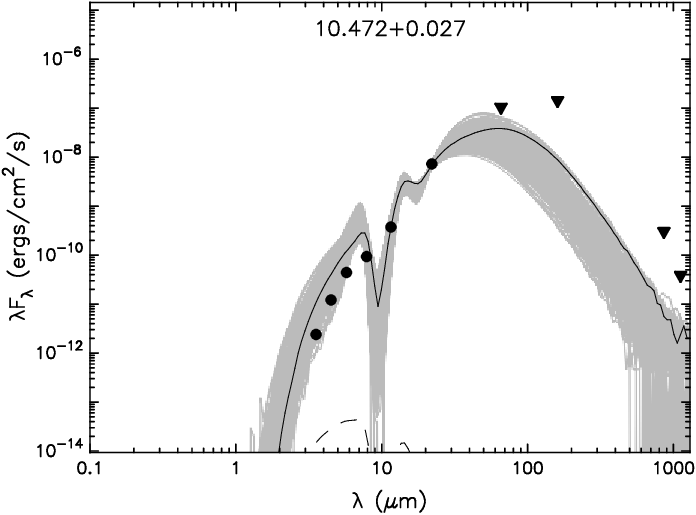}
\end{center}
\caption{The SED of G\,10.472+0.027 fitted by the \citet{Robitaille+07} online SED fitter.  The measurements (from left to right) are the GLIMPSE 3.6, 4.5, 5.8 and 8.0~\micron\/ {\em Spitzer} observations, the 12 and 22~\micron\/ WISE observations, the {\rm AKARI} 65 and 160~\micron\/ observations, the ATLASGAL 870~\micron\/ observation and the BOLOCAM Galactic Plane Survey 1.1~mm.  The solid black line shows the best fitting model, the gray lines are all models with a $\chi^2$ within a factor of 2 of the best fit model and the dashed line is the spectrum of the stellar photosphere.}
\label{fig:SED}
\end{figure}

The G\,10.472+0.027 masers are associated with an infrared point source detected in the GLIMPSE and WISE point source catalogues (source designations SSTGLMC G010.4723+00.0272 and J180838.23-195150.6, respectively) which are each offset by less than 1$^{\prime\prime}$ from the position we determined for the water maser (Fig.~\ref{fig:IR}).  Figure~\ref{fig:IR} shows that there are a number of additional bright, compact infrared sources close to the maser position and it is also embedded within a larger region of diffuse emission.  This means that observations of this region at angular resolutions coarser than a few arcseconds (i.e. most observations at far IR and sub-mm wavelengths) will likely contain contributions from more than one source.  We have used the measurements from the GLIMPSE I point source catalogue \citep{Churchwell+09}, the WISE All-Sky Source Catalog \citep{Wright+10}, the AKARI/FIS Bright Source Catalogue \citep{Yamamura+10}, the ATLASGAL Compact Source Catalogue \citep{Contreras+13} and the Bolocam Galactic Plane Survey (data release 2) \citep{Ginsburg+13} to estimate the spectral energy distribution (SED) of the high-mass star formation region (Fig.~\ref{fig:SED}).  Because of the lower angular resolution of the longer wavelength data (AKARI, ATLASGAL and Bolocam) these data have been considered as upper limits (emission was detected from each of these instruments at this location), as they almost certainly contain contributions from other nearby sources.  Indeed, comparison of the shorter wavelength WISE data with similar wavelength {\em Spitzer} data (which has higher angular resolution) shows significantly higher intensity in the former case.   The \citet{Robitaille+07} SED model fitter gives the mass of the central star as 19--49~\msol\/, (spectral type O9--O4) with an accretion rate of 0--$8.8 \times 10^{-3}~$\msol yr$^{-1}$ and a total luminosity of 47--290$\times 10^3$~\lsol\/ (we consider models with a $\chi^2$ within a factor of 2 of the best fit model as representing the parameter range consistent with the data).  \citet{Cesaroni+98} estimated a spectral type of B0 for the embedded high-mass star on the basis of radio continuum observations, however, they assumed a distance of 5.8~kpc in their calculations and correcting their luminosity for a distance of 8.55~kpc implies a spectral type of O9, consistent with the SED fitting.  The number of young, high-mass protostars contributing to the infrared through millimetre continuum used for the SED fitting, or the radio continuum observations is not presently known, but it appears very likely that there is more than one.  This suggests that the highest mass object in the vicinity of the water maser is either a late O-, or early B-star.  Further sensitive high resolution mid-infrared or sub-mm observations are require to identify the number of young high-mass protostars in the region and their location with respect to the water masers.

On the basis of the mass of the associated dust clump in the BGPS and ATLASGAL data G\,10.472+0.027 has recently been identified as a massive protostellar cluster, with a total cluster mass $\sim$25000--35000~\msol\/ (after correction to account for the more accurate distance) \citep{Ginsburg+12,Urquhart+13}.  Only the highest mass dust clumps are capable of hosting massive protostellar clusters and these evolve into massive stellar clusters such as NGC3603.  The BGPS and ATLASGAL surveys are sensitive to all such sources within in the solar circle and their results suggest that there are less than 20 such sources in the Galaxy \citep{Urquhart+13}.  A protostellar cluster of this mass will form 10's of O-stars.  The detection of the highest velocity water maser outflow within a massive protostellar cluster may be because this region contains one of only a handful of proto-O stars that exist in the Galaxy, or it may be that the extreme environment plays a critical role in producing an extreme outflow.  Higher spatial resolution observations of the water masers in this region may yield unique insights into the formation of stars in the the most massive clusters.

\section{Conclusions}

We have identified the highest velocity water maser emission associated with any outflow from a star formation region in the Milky Way.  The G\,10.472+0.027 water maser has a total velocity range of approximately 300~\kms\/, with the red-shifted emission offset from the systemic velocity of the region by 250~\kms.  The combination of wide velocity coverage with good spectral and high angular resolution, along with high sensitivity was crucial in allowing us to detect this extreme source.  The new generation of spectrometers available on interferometers such as the JVLA and ATCA will make it much easier to detect sources such as this in future.  Observations at mid-infrared through radio wavelengths are consistent with the water maser emission being associated with a late O- or early B-star embedded within a massive protostellar cluster.  Future high angular resolution observations of the masers offer the prospect of studying the kinematics of star formation within this extreme environment.

\section*{Acknowledgements}

This research has made use of NASAÕs Astrophysics Data System Abstract Service.  This research has made use of the NASA/ IPAC Infrared Science Archive, which is operated by the Jet Propulsion Laboratory, California Institute of Technology, under contract with the National Aeronautics and Space Administration.

\end{document}